\documentclass[reprint,amsmath,amssymb,aps,prb,showpacs]{revtex4-1}

\usepackage{graphicx}
\usepackage{dcolumn}
\usepackage{bm}

\usepackage{mathrsfs}
\usepackage{hyperref}
\usepackage{color}
\usepackage{soul}
\usepackage[normalem]{ulem}
\begin{document}
  
\title{ Revisiting the Hybrid  Quantum Monte Carlo Method for Hubbard and Electron-Phonon Models. }

\author{Stefan Beyl}
\author{Florian Goth}%
\author{Fakher F. Assaad}%

\affiliation{Institut f\"ur Theoretische Physik und Astrophysik,\\
Universit\"at W\"urzburg, Am Hubland, D-97074 W\"urzburg, Germany}

\date{\today}
\begin{abstract}
A unique feature of the hybrid quantum Monte Carlo (HQMC) method  is  the  potential to simulate  negative sign free lattice fermion models  with subcubic scaling in system size.  
Here we will revisit the algorithm for various models. We will show that for the  Hubbard model the HQMC  suffers from ergodicity issues and unbounded forces in the effective action. Solutions to these issues can be found   in terms of a complexification of the auxiliary fields. This implementation of the HQMC that does not attempt to regularize the  fermionic matrix so as to circumvent  the aforementioned singularities  does not  outperform   single spin flip  determinantal methods with cubic scaling.
On the other hand we will argue that there is a set of models for which the HQMC is very efficient. This class is characterized by effective actions free of singularities.
Using the Majorana representation, we show that  models such as the Su-Schrieffer-Heeger   Hamiltonian at half filling and on a bipartite lattice  belong to this class.
For this specific model sub-cubic scaling is achieved. 

\end{abstract}

\maketitle

\tableofcontents

\section{Introduction}
There has recently been  tremendous progress in  classifying  fermionic  model Hamiltonians that one can solve  with quantum Monte Carlo (QMC) methods without encountering the infamous negative sign problem \cite{Wu04,Yao14a,Li16,Wei16}.
Remarkably, this class of models  contains a number of extremely interesting  phases of matter  and quantum phase transitions \cite{Capponi00,Assaad04,Hohenadler10,Huffman14,Wang14a,Assaad16,Gazit16}.
Since in a lot of  these models the fermions are gapless, they cannot be integrated out, and are at the very origin of novel quantum critical phenomena \cite{Assaad13,Toldin14,Otsuka16,Berg12,Schattner15,Xu16c,SatoT17}.
From the technical point of view,  this leads to the necessity of developing  efficient  algorithms for lattice fermions  so as to unravel  facets of fermion critical phenomena in two and higher spacial dimensions. 

The condensed matter community has focused on the  Blankenbecler  Scalapino Sugar  (BSS)  auxiliary field determinantal QMC technique  \cite{Blankenbecler81,White89,Assaad08_rev}. This approach invariably scales as the cubed of the  volume since the fermionic determinant is explicitly calculated. Furthermore,  since the  determinant is nonlocal, cluster algorithms have remained elusive and single spin-flip updates are still the standard.   Using machine learning techniques to propose global moves is an ongoing active  research subject \cite{Xu17a,Wang17,Wang17b}.  On the other hand, in particle physics, especially in the lattice gauge theory community, the hybrid quantum Monte Carlo method (HQMC) was used and extended \cite{Scalettar87,Duane87,Kennedy2001}.   
A glimpse at the simulated system sizes fosters the hope to access larger lattice sizes, at least for a selected subset of models, by using HQMC. 
From a conceptual point of view, the HQMC method offers two main advantages in comparison to established methods in the condensed matter:
\begin{itemize}
\item A global updating scheme, based on Hamilton's equations of motion \cite{Kogut75}, that guarantees a good acceptance rate.
\item Replacing determinants by Gaussian integrals is the first key step to allow for subcubic scaling.   
\end{itemize}
Note that  global update schemes such as Langevin dynamics or hybrid Monte Carlo \cite{Kogut75}  can be implemented  in schemes that explicitly retain the fermion determinant. 

The structure and main results of this paper are as follows.
In Sec.~\ref{realAlog} we will revisit the ideas of  Ref.~\onlinecite{Scalettar87} as applied to the Hubbard model.
A number of subsections will  give us the opportunity to introduce notation and summarize important ideas   of the HQMC. 
In the final subsection, \ref{realResults},  we show that the algorithm is not ergodic, even at half filling when the weight is always positive.
The fermion determinant in each spin sector has a strongly fluctuating sign at low temperatures.  Particle-hole symmetry locks the relative signs of the determinants to unity and the weight is positive.   Nevertheless,  at  low temperatures the weight has many zero modes  and   different  regions of   configuration space are separated by divergences in the effective potential through which the molecular  dynamics cannot tunnel.  
In  Sec.~\ref{cHQMC} we  proceed 
to describe a complexification of the algorithm that circumvents these ergodicity issues.  In contrast to Ref.~\onlinecite{old_cHQMC}  our approach  is based on a complex Hubbard Stratonovich (HS) transformation. 
In Sec.  \ref{compareHQMCBSS}  we show that an ergodic algorithm can be achieved and that it can reproduce standard BSS results as obtained with the ALF package \cite{2017ALF}.
However, for the Hubbard model,  our implementation  of  the HQMC does not provide an improved scaling and is less efficient in terms of fluctuations. 
Here we note that  the BSS relies on a discrete Hubbard-Stratonovich transformation, thereby circumventing the above ergodicity issues. 

To proceed   we will ask the question if there is a class of models in the solid state for which the HQMC can be the method of choice.   To do so, we will follow the idea that the Hubbard model is hard, since the  effective action shows divergences and thereby generates forces that are unbounded.
With the help of recent progress in our understanding of the negative sign problem \cite{Yao14a,Li16,Wei16},
it can be shown that for a class of models the fermion determinant in a single spin sector is positive semidefinite.
This turns out to be the case for the so-called $n$-flavored Su-Schrieffer-Heeger (SSH)  model \cite{SSHmodel} at half filling  and on any bipartite lattice.
We have implemented the HQMC for this model and have benchmarked our results against the so-called continuous time interaction expansion (CT-INT) algorithm \cite{Rubtsov04,Assaad07,Assaad14_rev} where the phonons are integrated out  in favor of a retarded interaction.
It appears that this model can be efficiently   simulated with the HQMC.
For the comparison with other approaches, two points are in order. 
(i) There is by construction no   discrete-field  formulation of this model, and  local moves   lead to large autocorrelation times \cite{LangHohenadler2008}, and  
(ii) although very appealing, the CT-INT approach  suffers from a negative sign problem at finite phonon frequency and in dimensions larger than unity.
We will hence conclude  and give numerical evidence in Sec.~\ref{sec:HQMCSSH}  that the  HQMC could be the method of choice for this specific model Hamiltonian.
It is also interesting to note that the SSH model, in some limiting cases, maps onto a $Z_2$ lattice gauge theory as pointed out   in Ref.~\onlinecite{Assaad16}.
Finally  Sec.~\ref{Sec:conclusions}  draws some conclusions.

\section{HQMC for the Hubbard model}  \label{realAlog}
We start with a pedagogical introduction to the HQMC method in condensed matter
systems by revisiting Scalettar's\cite{Scalettar87} initial formulation.  This will provide us with the  
necessary background  to discuss failures and means to resolve them.
At the end of the section, we will discuss ergodicity issues.
\subsection{Basic formalism}\label{subsec:basic}
 The Hubbard Hamiltonian $\hat{H}$ is given as the sum of a kinetic part $\hat{H}_{\mathrm{K}}$ and an interacting part $\hat{H}_{\mathrm{U}}$,
\begin{equation}
 \hat{H}=\hat{H}_{\mathrm{K}}+\hat{H}_{\mathrm{U}}.
\end{equation}
While the kinetic energy is given by a tight binding Hamiltonian
\begin{equation}
 \hat{H}_{\mathrm{K}}=-t\sum_{\left\langle i,j\right\rangle ,\sigma}\left(\hat{c}_{i,\sigma}^{\dagger}\hat{c}_{j,\sigma}+\hat{c}_{j,\sigma}^{\dagger}\hat{c}_{i,\sigma}\right)
\end{equation}
and favors extended states, 
the potential energy is represented by an on-site Hubbard interaction 
\begin{equation}
 \hat{H}_{\mathrm{U}}=U\sum_{i}\left(\hat{n}_{i,\uparrow}-\frac{1}{2}\right)\!\left(\hat{n}_{i,\downarrow}-\frac{1}{2}\right),
\end{equation}
and favors localized states.   $\hat{c}^\dagger_{i,\sigma}$ ($\hat{c}^{\phantom{\dagger}}_{i,\sigma}$) denote  fermionic operators that create (annihilate) an electron  in a Wannier state centered around site  $i$ with a $z$ component of spin $\sigma$ and $\left\langle i,j\right\rangle$ denotes  nearest neighbors of a hyper cubic lattice.
The Hubbard interaction strength is given by $U$, $t$ denotes the hopping matrix element, and $\hat{n}_{i,\sigma} = \hat{c}^\dagger_{i, \sigma} \hat{c}^{\phantom{\dagger}}_{i, \sigma}$.
To describe thermodynamical properties, the partition function $Z$ is the quantity of interest. To compute it we  discretize the imaginary time $\tau$  and  introduce a Trotter decomposition
\begin{eqnarray}
 Z & = & \mathrm{tr}\:\mathrm{e}^{-\beta \hat{H}}\nonumber\\
 & = & \mathrm{tr}\left(\mathrm{e}^{-\triangle_{\tau}\hat{H}}\right)^{N_{\tau}}\nonumber\\
 & \simeq & \mathrm{tr}\left(\mathrm{e}^{-\triangle_{\tau}\hat{H}_{\mathrm{K}}}\mathrm{e}^{-\triangle_{\tau}\hat{H}_{\mathrm{U}}}\right)^{N_{\tau}}.
\end{eqnarray}
Here  $N_\tau \triangle_\tau = \beta$. 
The discretization into $\triangle_\tau$ slices is common to the HQMC and the BSS-QMC.   For the sake of comparison, it is important to note that both methods share the same $\triangle _\tau$ discretization error.
To be able to integrate out   the fermions, we have to decouple the many body interaction term into a  sum of single body propagators. This is achieved with a 
Hubbard-Stratonovich (HS) decomposition that introduces an auxiliary field $x_{i,l}$ at each site $i$ and every time slice $l$,
\begin{widetext}
 \begin{equation}
 \exp\left[-\triangle_{\tau}U\left(\hat{n}_{i,\uparrow}-\frac{1}{2}\right)\left(\hat{n}_{i,\downarrow}-\frac{1}{2}\right)\right]=\left(\triangle_{\tau}/\pi\right)^{1/2}\mathrm{e}^{-\triangle_{\tau}U/4}\intop_{-\infty}^{\infty}\mathrm{d}x_{i,l}\:\mathrm{ exp}\left\{ -\triangle_{\tau}\left[x_{i,l}^{2}+\sqrt{2U}x_{i,l}\left(\hat{n}_{i,\uparrow}-\hat{n}_{i,\downarrow}\right)\right]\right\}.  \label{HS_transformation}
 \end{equation}
\end{widetext}
In contrast to the discrete auxiliary field of the BSS algorithm \cite{Hirsch83,2017ALF}  this field is continuous. 
At  this point, we can integrate out the fermion degrees of freedom to obtain: 
\begin{equation}
 Z=\int\left[\delta x\right]\mathrm{ e}^{-S_{\mathrm{B}}(x)}\det M_{\uparrow}(x)\; \det M_{\downarrow}(x), \label{eq_introduceDets}
\end{equation}
where we have introduced the shorthand notation for the action of the auxiliary fields
\begin{equation}
 S_{\mathrm{B}}(x):=\triangle_{\tau}\sum_{i,l}x_{i,l}^{2}.
\end{equation}
The matrices $M_{\sigma}(x)$, appearing in the determinants, have a block structure:
\begin{equation}
 \!\! M_{\sigma}=\left(\!\begin{array}{cccccc}
I & 0 & 0 & \cdots & 0 & B_{N_{\tau},\sigma}\\
-B_{1,\sigma} & I & 0 & \cdots & 0 & 0\\
0 & -B_{2,\sigma} & I & \cdots & 0 & 0\\
\vdots & \vdots & \vdots & \ddots & \vdots & \vdots\\
0 & 0 & 0 & \cdots & I & 0\\
0 & 0 & 0 & \cdots & -B_{N_{\tau}-1,\sigma} & I
\end{array}\right)\!.
\label{B_Block}
\end{equation}
The dimension, $V_{S}\times V_{S}$, of the block matrices is determined by the number of sites $V_S$.
In the above,
\begin{equation}
 B_{l,\sigma}(x)=\mathrm{e}^{-\triangle_{\tau}K}\mathrm{e}^{-\sigma\triangle_{\tau}V_{l,\sigma}(x)} .
 \end{equation}
 $K$ represents the tight-binding hopping matrix with elements 
\begin{equation}
 K_{i,j}=\begin{cases}
-t & \left\langle i,j\right\rangle \mathrm{\,nearest\, neighbors}\\
0 & \mathrm{otherwise}
\end{cases},
\end{equation}
and the diagonal matrix $V$ contains the  fields
\begin{equation}
\left(V_{l}\right)_{i,j}(x)=\delta_{i,j}\sqrt{2U}x_{i,l}.
\end{equation}

Equation ~\eqref{eq_introduceDets} provides a suitable representation for discussing the absence of the fermionic sign problem.
For a half-filled bipartite lattice, where hopping occurs only between the two sublattices, it can be shown that both determinants have always the same sign since under a particle-hole transformation:
\begin{equation}
 \det M_{\downarrow}(x)=\mathrm{e}^{-\triangle_{\tau}\sqrt{2U}\sum_{i,l}x_{i,l}}\:\det M_{\uparrow}(x).
\end{equation}
Thus,  at half-filling the fermionic sign problem is absent, since all configurations of the auxiliary field have a positive statistical weight \cite{Hirsch85}.
One of the key steps  to avoid cubic scaling is to get rid of the determinant and to sample it stochastically.
To this aim, one introduces  so-called pseudo fermion fields
$\phi_{\sigma}$ to obtain
\begin{equation}
 \!Z=\!\int\!\!\left[\delta x\,\delta\phi_{\sigma}\right]\mathrm{e}^{-S_{\mathrm{B}}(x)-\sum_{\mathrm{\ensuremath{\sigma}}}\phi_{\sigma}^{\mathrm{T}}\left(M_{\sigma}^{\mathrm{T}}(x) M_{\sigma}(x)\right)^{-1}\phi_{\sigma}}.
 \label{eq:Zpseudofermion}
\end{equation}
Clearly the above implicitly assumes the absence of  negative sign problem, $ \det M_{\uparrow}(x)  \det M_{\downarrow}(x) > 0 $. 

\subsection{The Hybrid Monte Carlo updating scheme }\label{subsec:updates}
This subsection summarizes Hybrid Monte Carlo  sampling. 
Details on the implementation will be given at the end of this subsection.
In order to define a Hamiltonian system, we  add  a canonical conjugate variable, $p_{i,l}$,  to the HS field such that  Eq.~\eqref{eq:Zpseudofermion} reads:
\begin{equation}
 Z=\int\left[\delta x\,\delta p\,\delta \phi_{\sigma}\right]\: P(x, p, \phi_{\sigma})
 \label{eq:partitionfunction}
\end{equation}
with the distribution function
\begin{equation}
      P(x, p, \phi_{\sigma}) = \mathrm{e}^{-\mathcal{H}\left(x,p,\phi_{\sigma}\right)}
\end{equation}
and Hamiltonian
\begin{align}
 \mathcal{H}\left(x,p,\phi_{\sigma}\right)  := & \:S_{\mathrm{B}}(x)+\sum_{i,l}p_{i,l}^{2}+\label{eq:classHam}\\
   & +\sum_{\mathrm{\ensuremath{\sigma}}}\phi_{\sigma}^{\mathrm{T}}\left(M_{\sigma}^{\mathrm{T}}(x)\: M_{\sigma}(x)\right)^{-1}\phi_{\sigma}\nonumber.
\end{align}
We  now want to draw samples from the distribution $P(x,p, \phi_{\sigma})$ on the state space spanned by 
the set of continuous variables $\{x, p, \phi_{\sigma} \}$. The components of the momentum $p_{i,l}$ are distributed according to a Gaussian distribution and can be sampled directly.  
The auxiliary fields $\phi$ can be sampled in the following way.  Given the auxiliary variables
\begin{equation}
 R_{\sigma}:=\left(M_{\sigma}^{\mathrm{T}}(x)\right)^{-1}\phi_{\sigma},
\end{equation}
drawn from a Gaussian distribution, we can obtain $\phi$ from
\begin{equation}
 \phi_{\sigma}=M_{\sigma}^{\mathrm{T}}(x) R_{\sigma}.
\end{equation}
To sample the field  $x$  we use Hamilton's equations of motion  based on the Hamilton function  of Eq.~\eqref{eq:classHam} at fixed values of the pseudo fermion fields: 
\begin{align}
\dot{p}_{i,l} & =  -\frac{\partial\mathcal{H}}{\partial x_{i,l}}\\
\dot{x}_{i,l} & =  \frac{\partial\mathcal{H}}{\partial p_{i,l}}.
\end{align}
Integration over a given time interval  yields  a new point in phase space, $x'$ and $p'$, that we will accept according to the  Metropolis-Hastings rule:
\begin{equation}
\begin{split}
 r_{\mathrm{MH}} &= \min\left(1, \frac{T_0(x',p' \rightarrow  x,p)  }{ T_0(x,p \rightarrow  x',p')}\frac{P(x', p', \phi)}{P(x,p,\phi)}\right) \\
 & = \min\left(1, \mathrm{e}^{\mathcal{H}\left(x,p,\phi\right) - \mathcal{H}\left(x' ,p' ,\phi \right)}\right),
 \end{split}
\end{equation}
where $T_0$ corresponds to the proposal probability density. 
For the last identity  we have used the fact  that Hamiltonian dynamics is time reversal symmetric and  that  the phase space volume is conserved. Under these  assumptions, which will have to be satisfied by the numerical integrator (see below), $\frac{T_0(x',p' \rightarrow  x,p)  }{ T_0(x,p \rightarrow  x',p')} = 1$. 
Throwing a random number against $r_{\mathrm{MH}}$ we can then decide whether we accept  the update. Clearly, if the Hamiltonian propagation  is carried out exactly, the acceptance is unity. 
To conclude this overview we summarize the updating procedure
\begin{itemize}
 \item draw Gaussian samples for $p_{i,l}$ and $R_{\sigma,i,l}$.
 \item evolve $x$ and $p$ according to Hamilton's equations of motions.
 \item accept the new values of $x$ and $p$ according to the Metropolis-Hastings ratio.
\end{itemize}
\subsection{Detailed balance - the right choice of the integrator}\label{subsec:balance}
As alluded above, to integrate the equations of motion we have to proceed with care and remember that
Hamilton's equations of motion have two important properties: First, by Liouville's theorem the phase space volume is conserved, and second, they are time reversal symmetric.
Under the condition that we choose an integrator for Hamilton's equations of motions for the updates of $x$ and $p$ that retains these properties it follows that our Monte Carlo updates fulfill the detailed balance condition \cite{Liu2004a}.
Integrators that have these favorable traits are called symplectic, or geometric integrators \cite{Hairer2006} and the most well-known example of them is the Leapfrog method.
The Leapfrog method satisfies time reversibility and is well established in numerics. It solves the equation of motion iteratively over an artificial time.
The artificial Leapfrog time has to be discretized into time steps $\triangle t$.
Initially we propagate the momentum field by a half time step $\triangle t/2$. Afterwards we propagate alternatively the spatial field 
\begin{align}
 x_{i,l}\left(t+\triangle t\right) & -x_{i,l}\left(t\right) = \intop_{t}^{t+\triangle t}\text{d}t^{\prime}\;2p_{i,l}\left(t^{\prime}\right) \nonumber \\
 & =2p_{i,l}\left(t+\frac{1}{2}\triangle t\right)\triangle t+\mathcal{O}\left(\triangle t^{3}\right)
\end{align}
and the momentum field 
\begin{align}
 p_{i,l} & \left(t+\frac{3\triangle t}{2}\right) -p_{i,l}\left(t+\frac{\triangle t}{2}\right) \nonumber \\
 & = -\intop_{t+1/2\,\triangle t}^{t+3/2\,\triangle t}\text{d}t^{\prime}\;\frac{\partial\mathcal{H}\left(x,t^{\prime}+\triangle t\right)}{\partial x_{i,l}} \nonumber \\
 & =-\frac{\partial\mathcal{H}\left(x,t+\triangle t\right)}{\partial x_{i,l}}\triangle t+\mathcal{O}\left(\triangle t^{3}\right)
\end{align}
until we reach the stopping time, for example $1/\triangle t$.
The Leapfrog method, as well as the conjugate gradient method (see below) has some systematical errors.
To ensure their controllability at the end of each Leapfrog run we perform a Metropolis check to decide if we accept or reject the new auxiliary field configuration.
It turns out that the acceptance depends on the size of the artificial Leapfrog time steps $\triangle t$ as well as on the accuracy of the conjugate gradient method as studied in Ref.~\onlinecite{Kennedy2001}.
Our experience tells us, that configurations will be rejected if the system is in a point of the phase space where the product of the gradient of the potential times the velocity in the according direction is large, compared to the artificial time steps.
Therefore a second Leapfrog run with the same auxiliary field configuration but different initial momentum field may have much better propensity to produce a new configuration that will be accepted. If several Leapfrog runs fail, the time steps have to be made smaller. 

We also implemented an adaptive Leapfrog method\cite{Hairer2005} that is also time reversible but selects the step-size according to the gradient of the Hamiltonian, $\triangle t \propto 1/ |\nabla H |^2$.
Summarizing, the acceptance is very good, but if a single bad conditioned configuration occurs, it slows down to very small time steps by trying to solve the equations of motions in regions of high variability and needs a very long time.
These regions very likely correspond to regions discussed in Sec. \ref{realResults} where the determinants change signs and therefore the symplectic integrator has tried to accommodate for the singularity in the gradient.
Leapfrogs with fixed step size may fail at this configuration for several times, but after several runs with different momentum configurations they also will find a configuration that will be accepted, if the selected time step is not too large.
Another way, we observed, to improve the acceptance is to substitute a Leapfrog run by several shorter runs. This generates the side effect of reduced correlations between measurements in some cases.
We attribute this behavior to the additional generation of random $\phi_{\sigma}$ and $p_{\sigma}$ fields between measurements.
For larger system sizes or higher values of $\beta$ it can be necessary to shorten the time steps $\triangle t$ to keep the acceptance high.
\subsection{Evaluation of the forces and measurement of observables}\label{subsec:observables}
The evaluation of derivative, which is necessary to calculate the forces during the Leapfrog simulation, is simplified by using an algebraic identity as well as the symmetry of $M_{\sigma}^{\mathrm{T}}M_{\sigma}$ to obtain
\begin{widetext}
 \begin{equation}
 \frac{\partial}{\partial x_{i,l}}\phi^{\mathrm{T}}_{\sigma}\left(M_{\sigma}^{\mathrm{T}}\left(x\right)M_{\sigma}\left(x\right)\right)^{-1}\phi_{\sigma}=
 -2\phi^{\mathrm{T}}_{\sigma}\left(M_{\sigma}^{\mathrm{T}}\left(x\right)M_{\sigma}\left(x\right)\right)^{-1}\left[M_{\sigma}^{\mathrm{T}}\left(x\right)\left(\frac{\partial}{\partial x_{i,l}}M_{\sigma}\left(x\right)\right)\right]\left(M_{\sigma}^{\mathrm{T}}\left(x\right)M_{\sigma}\left(x\right)\right)^{-1}\phi_{\sigma}.
\end{equation}
\end{widetext}
Measurements can be performed for every new configuration of the auxiliary fields.
The bare one particle Green's function is given by the inverse $M_{\sigma}$ matrix.
Instead of a, numerically expensive, inversion of $M_{\sigma}$, every component $(M_{\sigma}^{-1})_{i,j}$ can be sampled by using $2[(M_{\sigma}^{\mathrm{T}}  M_{\sigma})^{-1}\phi_{\sigma}]_{i}[R_{\sigma}]_{j}$ as an unbiased estimator.
Wick's theorem allows us to calculate also many particle observables.
An efficient method for dealing with the linear system 
\begin{equation}
 M_{\sigma}^{\mathrm{T}}M_{\sigma} X = \phi  \label{eq:CGmotivation}
\end{equation}
and we find the solution vector $X$ is of utmost importance for an efficient implementation and is required for the sampling procedure in every Leapfrog step as well as for the measurement of observables.
A more detailed discussion how to solve those systems of linear equations is discussed in the next subsection.

\subsection{Solution of linear systems with the conjugate gradient method}\label{realCG}
In favor of readability we will suppress the spin index $\sigma$ in this subsection, since the spin up and spin down sectors are treated in exactly the same way.
As is well known, the straight forward inversion of an $N\times N$ matrix by basic Gaussian elimination needs $\mathcal{O}(N^{3})$ flops.
As mentioned above, we only need to know the solution of the linear system \eqref{eq:CGmotivation} to formulate the complete algorithm.
In contrast to the inversion of a matrix, iteratively finding the solution to a system of linear equations up to the computing precision
can be possible in an amount of computing time that is linear in the entries of the matrix and hence we can benefit greatly, if the system has a sparse  system matrix.
The workhorse method for symmetric, linear systems is the conjugate gradient(CG) method
which we will explain on the basis of the prototypical system
\begin{equation}
 O X=\phi. \label{eq_CGinit}
\end{equation}
The matrix $O = M^{\mathrm{T}}M$, is symmetric and positive semidefinite.
The solution to Eq.~\eqref{eq_CGinit} can be interpreted as the unique minimum of the quadratic function
\begin{equation}
 f(X) = \frac{1}{2} X^T O X - X^T \phi.
\end{equation}
Given a starting guess $X_0$, the idea is now not to take the direct gradient evaluated at each step but to choose search directions that are orthogonal, with respect to the from $O$ induced scalar product, to all previously constructed directions.
Employing that idea the following iterative prescription emerges.
\begin{align}
 & X_{n+1}=X_{n}+\zeta_{n}d_{n} \nonumber\\
 & r_{n+1} = r_{n}-\zeta_{n}Od_{n}\nonumber\\
 & \eta_{n}= \langle r_{n+1},r_{n+1}\rangle/\langle r_{n},r_{n} \rangle \\
 & d_{n+1}=r_{n+1}+\eta_{n}d_{n} \nonumber\\
 & \zeta_{n+1}=\langle r_{n+1},r_{n+1}\rangle/\langle d_{n+1},Od_{n+1}\rangle \nonumber
\end{align}
with iterative approximations $X_n$ to the true solution $X$.
We define the residual after the $n$th iteration by
\begin{equation}
 r_{n}=\phi -O X_{n}.
\end{equation}
The absolute value of the residual vector is something like a measure of how close the approximated solution is to the exact solution. Therefore, we use it to define a termination criterion for the iterative conjugate gradient method
\begin{equation}
 \varepsilon_n = \sqrt{\frac{\left(\phi-O X_n \right)^{2}}{\phi^{2}}}\leq10^{-7},
\end{equation}
similar to the criterion chosen in Ref.~\onlinecite{Scalettar87}.
The notation $\langle r,d \rangle$ represents a scalar product between two vectors $r$ and $d$.
To start the iterative procedure we have to choose an initial vector $X_{0}$.
If nothing is known about the system of equations it can be any arbitrary vector, like a vector consisting of zeros.
A well designed guess can speed up the method and lower the number of iterations until it converges.
Given $X_0$ the initial data is completed by
\begin{align}
 &r_{0}=\phi -OX_{0}\nonumber\\
 &d_{0}=r_{0}.
\end{align}
In general, it is possible to show that
\begin{equation}
 \langle d_{m},Od_{n}\rangle = 0 \quad \mathrm{for }\quad m \neq n.
\end{equation}
Therefore, all $d_{n}$ vectors are linearly independent, with respect to $O$,  and a calculation with exact arithmetic would deliver the exact result after $N$ iterations.
Like the authors of Ref.~\onlinecite{Scalettar87} mention, it is very common to speed up a conjugate gradient algorithm by introducing a preconditioner.
It helps especially if the matrix is ill conditioned, which is known to be the case 
for stronger interaction strengths $U$ and larger values of $\beta$ \cite{BaiPropertiesofHubbardMatrices}.
To define a suitable preconditioner we need a matrix $\tilde{O}$ that is close to the matrix, representing the system of linear equations, but easy to invert.
Because the matrix $O$ is symmetric and positive semi definite, a good starting point is to use the Cholesky decomposition of $O$,
\begin{equation}
 \tilde{O}=L^{\mathrm{T}}L.
\end{equation}
The matrix $L$ is a triangular matrix and thus easy to invert.
We rewrite Eq.~\eqref{eq_CGinit}
\begin{equation}
 O^{\prime}X^{\prime}=\phi^{\prime}, \label{eq_CGinitMOD}
\end{equation}
and we substitute
\begin{align}
 &O^{\prime}=L^{-\mathrm{T}}OL^{-1}\nonumber\\
 &X^{\prime}=LX\\
 &\phi^{\prime}=L^{-\mathrm{T}}\phi,\nonumber
\end{align}
from which it follows that
\begin{equation}
 r^{\prime}_{n}=\phi^{\prime}-O^{\prime}X^{\prime}_{n}=L^{-\mathrm{T}}\,r_{n}.
\end{equation}
According to its definition, $O^{\prime}$ is also a symmetric and positive semi definite matrix.
Therefore, we can apply the conjugate gradient method to Eq.~\eqref{eq_CGinitMOD} to solve the modified system of linear equations. If we modify the iteration scheme, we get $X$ out of
\begin{align}
 & X_{n+1}=X_{n}+\zeta^{\prime}_{n}d_{n} \nonumber\\
 & r_{n+1} = r_{n}-\zeta^{\prime}_{n}Od_{n}\nonumber\\
 & \eta^{\prime}_{n}= \langle r_{n+1},\tilde{O}^{-1}r_{n+1}\rangle/\langle r_{n},\tilde{O}^{-1}r_{n}\rangle \\
 & d_{n+1}=r_{n+1}+\eta^{\prime}_{n}d_{n} \nonumber\\
 & \zeta^{\prime}_{n+1}=\langle r_{n+1},\tilde{O}^{-1}r_{n+1}\rangle/ \langle d_{n+1},Od_{n+1}\rangle. \nonumber
\end{align}
Once again, the numerical effort denies us the use of a usual Cholesky decomposition. Its exact calculation would be equivalent to the inversion of the matrix $O$.
Instead, Ref.~\onlinecite{Scalettar87} proposes an incomplete Cholesky decomposition.
They use the matrix product $M^{\mathrm{T}}M$, without hopping interactions and a slight shift of the diagonal elements. The matrix we decompose is given by the matrix of Eq.~\eqref{eq_O0}.
\begin{widetext}
 \begin{equation}
 \tilde{O}_{0}=\left(\begin{array}{cccccc}
\alpha I+B_{0,1}^{2} & -B_{0,1} & 0 & \cdots & 0 & B_{0,N_{\tau}}\\
-B_{0,1} & \alpha I+B_{0,2}^{2} & -B_{0,2} & \cdots & 0 & 0\\
0 & -B_{0,2} & \alpha I+B_{0,3}^{2} & \cdots & 0 & 0\\
\vdots & \vdots & \vdots & \ddots & \vdots & \vdots\\
0 & 0 & 0 & \cdots & \alpha I+B_{0,N_{\tau}-1}^{2} & -B_{0,N_{\tau}-1}\\
B_{0,N_{\tau}} & 0 & 0 & \cdots & -B_{0,N_{\tau}-1} & \alpha I+B_{0,N_{\tau}}^{2}
\end{array}\right) \label{eq_O0}
\end{equation}
\end{widetext}
All block matrices $B_{0,i}$ are now diagonal. The slight shift, e.g. $\alpha=1.05$, prevents the matrix to become ill conditioned and prevents a pivot breakdown of the CG \cite{PreconPaper}.
The matrix defined in Eq.~\eqref{eq_O0} can not be inverted analytically for $\alpha \neq 1$ \cite{Scalettar87}.
The new preconditioner matrix is given by
\begin{equation}
 \tilde{O}^{\prime}=L^{\mathrm{T}} D L.
\end{equation}
With diagonal matrix
\begin{equation}
 D=\left(\begin{array}{cccccc}
D_{1} & 0 & 0 & \cdots & 0 & 0\\
0 & D_{2} & 0 & \cdots & 0 & 0\\
0 & 0 & D_{3} & \cdots & 0 & 0\\
\vdots & \vdots & \vdots & \ddots & \vdots & \vdots\\
0 & 0 & 0 & \cdots & D_{N_{\tau}-1} & 0\\
0 & 0 & 0 & \cdots & 0 & D_{N_{\tau}}
\end{array}\right),
\end{equation}
as well as the triangular matrices
\begin{equation}
 L=\left(\begin{array}{cccccc}
I & -L_{1} & 0 & \cdots & 0 & L_{N_{\tau}}\\
0 & I & -L_{2} & \cdots & 0 & 0\\
0 & 0 & I & \cdots & 0 & 0\\
\vdots & \vdots & \vdots & \ddots & \vdots & \vdots\\
0 & 0 & 0 & \cdots & I & -L_{N_{\tau}-1}\\
0 & 0 & 0 & \cdots & 0 & I
\end{array}\right).
\end{equation}
In the atomic limit now all of them consist of diagonal block matrices of size $V_{S}\times V_{S}$.
Recursive definitions of those matrices keep the numerical effort down. The matrices for $l=1,\dots,N_{\tau}-1$ are given by
\begin{align}
 &L_{l}=D_{l}^{-1}B_{0,l}\\
 &D_{l}=\alpha I+B_{0,l}^{2}-B_{0,l-1}D_{l-1}^{-1}B_{0,l-1},\nonumber
\end{align}
while for $l=N_{\tau}$
\begin{align}
  L_{N_{\tau}}=&D_{1}B_{0,N_{\tau}}\\
 D_{N_{\tau}}=&\alpha I+B_{0,N_{\tau}}^{2}-B_{0,N_{\tau}-1}D_{N_{\tau}-1}^{-1}B_{0,N_{\tau}-1},\nonumber\\
 & -B_{0,N_{\tau}}D_{1}^{-1}B_{0,N_{\tau}}\nonumber
\end{align}
With these preliminaries the preconditioned conjugate gradient method can be used as described above.
One could ask the question, if the preconditioner we use is a good choice for our purposes.
It is not easy to answer this question precisely.
We are following the arguments of Scalettar \textit{et~al.}~in Ref.~\onlinecite{Scalettar87}, where they are proposing that the condition of the matrix,
and thus the number of conjugate gradient iterations, depends essentially on the interaction strength $U$.
The search for a better preconditioner is still ongoing and the only progress we know of can be found in Ref. \onlinecite{PreconPaper}, where progress for very strong
interactions is reported, although we like to point out that their tests were not performed on real configurations from a full Monte Carlo simulation but on randomly generated configurations of the auxiliary field.
The preconditioner we have chosen is well suited for interactions that are not too strong and especially fast to calculate because of the sparsity structure of our matrix.
Nevertheless even the preconditioned system can come to a configuration where the CG method never reaches the claimed accuracy; hence we will stop it after $N$ iterations and take the solution vector of this iteration or the one with the smallest deviation.
In general this problem occurs only for simulations with very inappropriate parameters and can be absorbed in the considerations of acceptance rates below.

\subsection{Ergodicity issues}\label{realResults}
\begin{figure}
 \includegraphics[width=.45\textwidth]{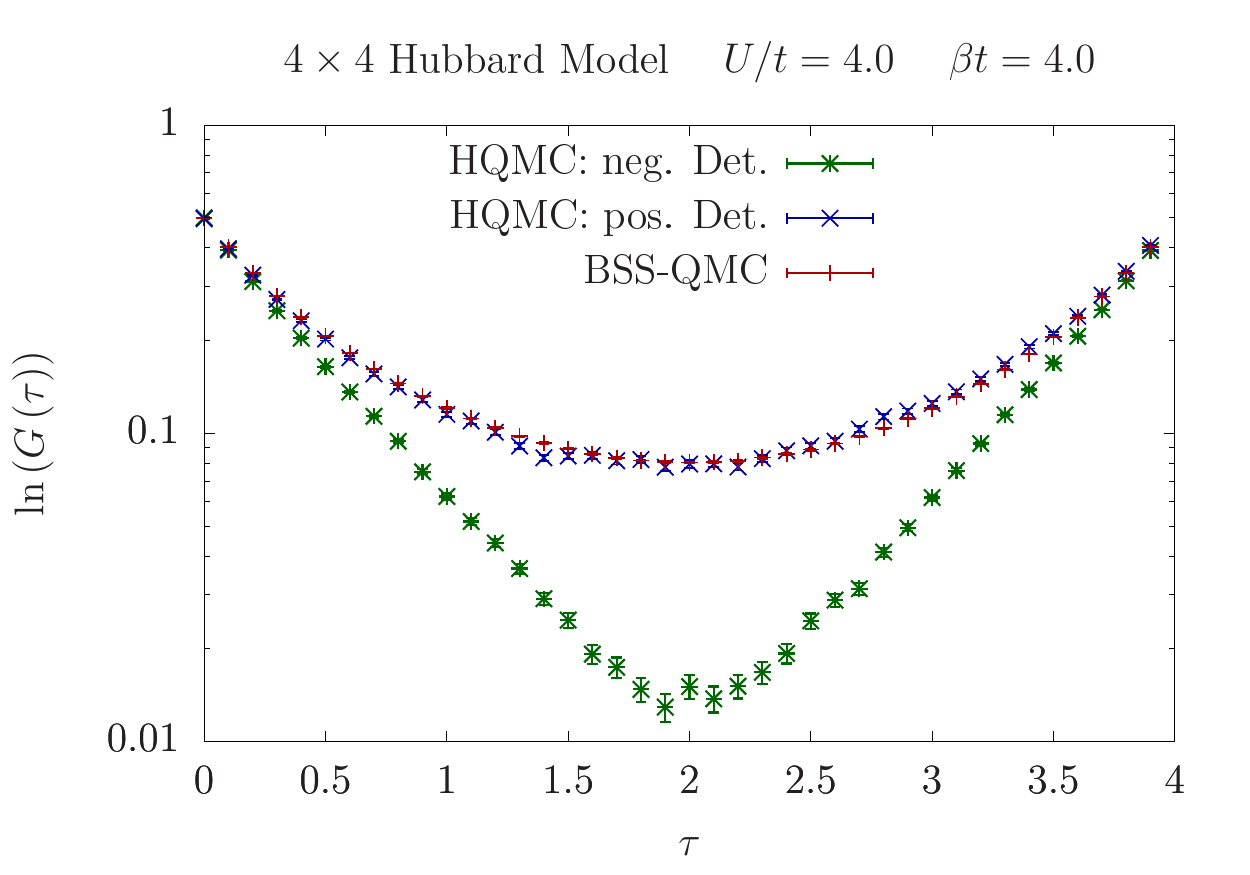}
 \caption{\label{fig_compare} Green's functions, calculated with the hybrid quantum Monte Carlo and the BSS-QMC \cite{2017ALF} algorithms  on a $4 \times 4$-lattice at $U=4$ and $\triangle_{\tau}=0.1$.
 Starting the simulation in different sectors clearly produces different results. }
\end{figure}
The implementation described until now, was already discussed in the original paper \cite{Scalettar87}.
We were able to reimplement the method and run it with several lattice configurations. Furthermore we also observed that for small values of $\beta$, the configuration space is strongly dominated by configurations where both determinants are positive and hence the Monte Carlo process encounters no difficulties in sampling this wide space.
If we increase the inverse temperature, the configuration space becomes much more fissured into  $(+,+)$   ( $ \text{sign}\;{\det(M_{\sigma})}  = 1$ ) and 
$(-,-)$  ($  \text{sign} \; {\det(M_{\sigma})}  = -1$)  domains.
Since the determinants are real continuous functions of a continuous field,   $(+,+)$   and $(-,-)$  domains are  necessarily separated by regions where 
$ \left(M_{\sigma}^{\mathrm{T}}  M_{\sigma}\right)^{-1}  $ diverges. 

This clearly poses a problem, since  any implementation of molecular dynamics with symplectic integrators  conserves  energy and will 
in principle never cross such  a barrier\cite{Barriers}. 
In particular, every molecular dynamics run will be trapped in the pocket of the configuration space where it started.
Since  the simple sampling of the momenta and pseudofermion  fields does not allow us to cross the barriers, a  Monte Carlo run will not be ergodic and can converge to a wrong result as demonstrated in Fig.~\ref{fig_compare}.
The figure demonstrates that starting a run in a $(+,+)$ or $(-,-)$  domain indeed leads to different results. 

The behavior of different integrators is quite interesting close to those boundaries. 
If we use the simple Leapfrog with large step sizes, it is able to cross over into other domains
by violating energy conservation.
The adaptive integrator \cite{Hairer2005} on the other hand will in this setting dutifully correct its step size down to vanishingly small values to accommodate for the large gradient of the potential that it tries to sample.
It will not cross the boundary but it will slow down the simulation indefinitely while trying to integrate the equations of motion with and preserving the total energy conservation.

\section{Complex reformulation and Comparison}\label{cHQMC}
One can follow several routes to circumvent the ergodicity issues of the algorithm using real auxiliary fields.
On one hand one could try to eliminate the potential barriers in  configuration space by a suitable transformation of the  problem.  Another idea is to modify the Leapfrog so that it can explicitly  tunnel between the barriers \cite{Barriers}. 
Finally  combining importance sampling schemes such as in the BSS algorithm  with the HQMC has been proposed in Ref.~\onlinecite{Imada88}   and has the potential to circumvent ergodicity issues.
Here we will follow the idea  of extending the configuration space to complex numbers \cite{old_cHQMC}.
\subsection{Complexification}
In a complex configuration space the barriers still exist but, due to  the additional degrees of freedom, the integrator should be able to produce trajectories that move around them.
The determinants become complex numbers and are conjugate to each other, thereby ensuring the absence of a negative sign problem. 
To achieve this complexification,  we  decouple the Hubbard interaction in charge and spin sectors by introducing a free parameter  $\alpha$: 
\begin{align}
 H_{V}'  = & \sum_{i}\alpha\frac{U}{2}\left(\hat{n}_{i,\uparrow}+\hat{n}_{i,\downarrow}-1\right)^{2} \nonumber \\
   & -\sum_{i}\left(1-\alpha\right)\frac{U}{2}\left(\hat{n}_{i,\uparrow}-\hat{n}_{i,\downarrow}\right)^{2}.
\end{align}
 $\alpha$ can be chosen from $[0, 1]$ thereby interpolating between the purely real code at $\alpha=0$ and a purely imaginary one at $\alpha = 1$.
Clearly the final result will be $\alpha$ independent, but it  will   determine the Monte Carlo configuration space.  In this section, we will redefine some of the symbols, variables and matrices that we have introduced previously.
At $\alpha = 1$   we can make contact with the formulation of Refs.~\onlinecite{Brower12,Buividovich12,Ulybyshev2013,Hohenadler14,Smith14,Luu16}   where a purely imaginary field couples to the local density.   In this case, and as argued in Ref.~\onlinecite{Ulybyshev17}  one equally  expects to encounter ergodicity issues.

To distinguish between the original and the new definitions, all redefined variables will be labeled by a prime.
The new formulation leads to a  doubling   of auxiliary fields:
\begin{widetext}
\begin{align}
 \mathrm{e}^{-\triangle_{\tau}H_{\mathrm{U}}^{\prime}}\propto\int \mathrm{d}x_{i,l}\: \mathrm{d}y_{i,l} & \:\exp\left\{ -\triangle_{\tau}\sum_{i}\left[x_{i,l}^{2}+y_{i,l}^{2}+\left(\sqrt{2U\left(1-\alpha\right)}x_{i,l}+\mathrm{i}\sqrt{2U\alpha}y_{i,l}\right)\left(\hat{n}_{i,\uparrow}-\frac{1}{2}\right)\right.\right.\nonumber\\
 & \left.\left.-\left(\sqrt{2U\left(1-\alpha\right)}x_{i,l}-\mathrm{i}\sqrt{2U\alpha}y_{i,l}\right)\left(\hat{n}_{i,\downarrow}-\frac{1}{2}\right)\right]\right\} ,
\end{align} 
\end{widetext}
and we also redefine 
\begin{equation}
 S_{\mathrm{B}}^{\prime}\left(x,y\right):=\triangle_{\tau}\sum_{i}\!\left(x_{i}^{2}+y_{i}^{2}\right).
\end{equation}
To shorten the notation, we will sometimes use a combined notation and interpret both auxiliary fields as components of one complex field
\begin{equation}
 z_{i,l}:=\sqrt{2U\left(1-\alpha\right)}\:x_{i,l}+\mathrm{i}\sqrt{2U\alpha}\:y_{i,l}.
\end{equation}
We formulate the partition function for complex auxiliary fields similar to the real case
\begin{align}
 Z  = & \int\left[\delta z\:\delta\overline{z}\right]\:\mathrm{e}^{-S_{\mathrm{B}}\left(z,\overline{z}\right)}\:\mathrm{e}^{\triangle_{\tau}\sum_{i,l}z_{i,l}/2}\det M_{\uparrow}^{\prime}\left(z,\overline{z}\right)\times\nonumber\\
   & \qquad\times\:\mathrm{e}^{-\triangle_{\tau}\sum_{i,l}\overline{z_{i,l}}/2}\det M_{\downarrow}^{\prime}\left(z,\overline{z}\right).
\end{align}
The matrices for the spin up and spin down sector still have the same structure as before
\begin{equation}
\!M_{\sigma}^{\prime}\!=\!\!\left(\begin{array}{cccccc}
I & 0 & 0 & \cdots & 0 & B_{N_{\tau},\sigma}^{\prime}\\
-B_{1,\sigma}^{\prime} & I & 0 & \cdots & 0 & 0\\
0 & -B_{2,\sigma}^{\prime} & I & \cdots & 0 & 0\\
\vdots & \vdots & \vdots & \ddots & \vdots & \vdots\\
0 & 0 & 0 & \cdots & I & 0\\
0 & 0 & 0 & \cdots & -B_{N_{\tau}-1,\sigma}^{\prime} & I
\end{array}\right)\!,\!\!
\end{equation}
with 
\begin{equation}
 B_{l,\sigma}^{\prime}\left(z,\overline{z}\right)=\mathrm{e}^{-\triangle_{\tau}K}\mathrm{e}^{-\sigma\triangle_{\tau}V_{l,\sigma}^{\prime}\left(z,\overline{z}\right)},
\end{equation}
and
\begin{equation}
 \left(V_{l,\sigma}^{\prime}\right)_{i,j}\left(z,\overline{z}\right)=\begin{cases}
z_{i,l}\delta_{i,j} & \sigma=\uparrow\\
\overline{z_{i,l}}\delta_{i,j} & \sigma=\downarrow
\end{cases}.
\end{equation}
At half filling, particle-hole symmetry induces a relation between the determinants of the spin matrices and their prefactors
\begin{align}
 \mathrm{e}^{-\triangle_{\tau}\sum_{i,l}\frac{\overline{z_{i,l}}}{2}}\det M_{\downarrow}^{\prime} = & \overline{\mathrm{e}^{\triangle_{\tau}\sum_{i,l}\frac{z_{i,l}}{2}}\det M_{\uparrow}^{\prime}}\nonumber\\
  = & \mathrm{e}^{\triangle_{\tau}\sum_{i,l}\frac{\overline{z_{i,l}}}{2}}\det M_{\uparrow}^{\prime\dagger}.
\end{align}
This means, that the statistical weight for every configuration is positive such that complexification does not lead to a sign problem.
Furthermore, we can use this relation to simplify the algorithm by substituting the spin down sector by the complex conjugate of the spin up sector
\begin{alignat}{2}
 Z & =  \int\left[\delta z\:\delta\overline{z}\right]\:&&\mathrm{e}^{-S_{\mathrm{B}}^{\prime}\left(z,\overline{z}\right)}\mathrm{e}^{\triangle_{\tau}\sum_{i,l}\frac{z_{i,l}+\overline{z_{i,l}}}{2}}\times\nonumber\\
 &  \qquad\qquad &&\times\det M_{\uparrow}^{\prime}\left(z,\overline{z}\right)\det M_{\uparrow}^{\prime\dagger}\left(z,\overline{z}\right)\nonumber\\
 & =  \int\left[\delta x\:\delta y\right]\: && \mathrm{e}^{-S_{\mathrm{B}}^{\prime}\left(x,y\right)}\det\left[\mathrm{e}^{\triangle_{\tau}\frac{\sqrt{2U\left(1-\alpha\right)}}{N_{\tau}V_{S}}\sum_{i,l}x_{i,l}}\times\right.\nonumber\\
 &   \qquad\qquad &&\left.\times M_{\uparrow}^{\prime}\left(x,y\right)M_{\uparrow}^{\prime\dagger}\left(x,y\right)\right]\nonumber\\
 & =  \int\left[\delta x\:\delta y\right] && \mathrm{e}^{-S_{\mathrm{B}}^{\prime}\left(x,y\right)}\det\left[\mathcal{M_{\uparrow}}\left(x,y\right)\mathcal{M_{\uparrow}^{\dagger}}\left(x,y\right)\right].
\end{alignat}
Here we defined
\begin{equation}
 \mathcal{M_{\uparrow}}\left(x,y\right):=\kappa(x) M_{\uparrow}^{\prime}\left(x,y\right)
\end{equation}
and
\begin{equation}
 \kappa(x):=\exp\left\{ \triangle_{\tau}\frac{\sqrt{2U\left(1-\alpha\right)}}{2N_{\tau}V_{S}}\sum_{i,l}x_{i,l}\right\}.
\end{equation}
As in Sec. \ref{realAlog}, the determinants can be eliminated by the use of a Gaussian identity, which introduces additional complex fields
\begin{equation}
 Z=\int\left[\delta x\:\delta y\:\delta\phi^{\prime}_{\uparrow}\:\delta\phi^{\prime\dagger}_{\uparrow}\right]\:\mathrm{e}^{-S_{\mathrm{B}}^{\prime}-\phi^{\prime\dagger}_{\uparrow}\left(\mathcal{M}_{\uparrow}^{\dagger}\:\mathcal{M}_{\uparrow}\right)^{-1}\phi^{\prime}_{\uparrow}}.
\end{equation}
The  partition function is now comparable in its form to the previous real algorithm. 
Instead of a real system of linear equations we now have to solve a complex one.
Although the number of degrees of freedom is  twice as large, we can use a complex version of the conjugate gradient method. Since we eliminated the spin down sector in the formulation, we get away with one call to the CG method.
The sampling of random numbers works analogously to the real version of the algorithm. We redefine
\begin{equation}
 R_{\sigma}^{\prime}:=\left(\mathcal{M}_{\sigma}^{\mathrm{T}}(x,y)\right)^{-1}\phi_{\sigma}^{\prime}\;\Rightarrow Z\propto\mathrm{e}^{-R_{\sigma}^{\prime\dagger}R^{\prime}_{\sigma}},
\end{equation}
and generate Gaussian random numbers for the real and imaginary part of $R^{\prime}_{\sigma}$ and perform a matrix vector multiplication to get the complex $\phi_{\sigma}^\prime$ fields
\begin{equation}
\phi_{\sigma}^{\prime}=\mathcal{M}_{\sigma}^{\dagger}(x,y) R_{\sigma}^{\prime}.
\end{equation}
Since we doubled our auxiliary field components, we also have to introduce two momentum fields to get a new artificial Hamiltonian 
\begin{align}
 \mathcal{H}^{\prime}&\left(x,y,p_{x},p_{y},\phi_{\uparrow}^{\prime},\phi_{\uparrow}^{\prime\dagger}\right)  := \nonumber\\
 & S_{\mathrm{B}}^{\prime}(x,y)+\sum_{i,l}\left(\left(p_{x}\right)_{i,l}^{2}+\left(p_{y}\right)_{i,l}^{2}\right)+\\
   &\qquad +\phi_{\uparrow}^{\prime\dagger}\left(\mathcal{M}_{\uparrow}^{\dagger}(x,y)\: \mathcal{M}_{\uparrow}(x,y)\right)^{-1}\phi_{\uparrow}^{\prime}\nonumber.
\end{align}
Both momentum fields are Gaussian distributed and thereby easy to sample. Hamilton's equations of motion lead the way to the complex Leapfrog method. 
\begin{align}
 & \dot{x}_{i,l}=\frac{\partial \mathcal{H}^{\prime}(x,y)}{\partial(p_{x})_{i,l}} =2(p_{x})_{i,l},\nonumber\\
 & \dot{y}_{i,l}=\frac{\partial \mathcal{H}^{\prime}(x,y)}{\partial(p_{y})_{i,l}} =2(p_{y})_{i,l},
\end{align}
\begin{equation}
 (\dot{p}_{x})_{i,l}=-\frac{\partial \mathcal{H}^{\prime}(x,y)}{\partial(x)_{i,l}},\qquad (\dot{p}_{y})_{i,l}=-\frac{\partial \mathcal{H}^{\prime}(x,y)}{\partial(y)_{i,l}}.
\end{equation}
Since the equations of motion decouple for the two parts of the auxiliary field, we find that the complex version of the algorithm just leads to a doubling of the degrees of freedom.
To measure the single  particle Green's function we invert the $M_{\sigma}^{\prime}$ matrix stochastically and come to the unbiased estimator $[(M_{\sigma}^{\prime\dagger}  M_{\sigma}^{\prime})^{-1}\phi_{\sigma}]_{i}[R_{\sigma}^{\prime}]_{j}$ for $(M_{\sigma}^{\prime -1})_{i,j}$.
Since we are dealing with complex numbers, also the estimator will give complex estimates.
In the process of averaging the observables over the Monte Carlo configurations their imaginary part has to vanish.
\subsection{Results of the complex method\label{complexResults}}
\begin{figure}
 \includegraphics[width=.45\textwidth]{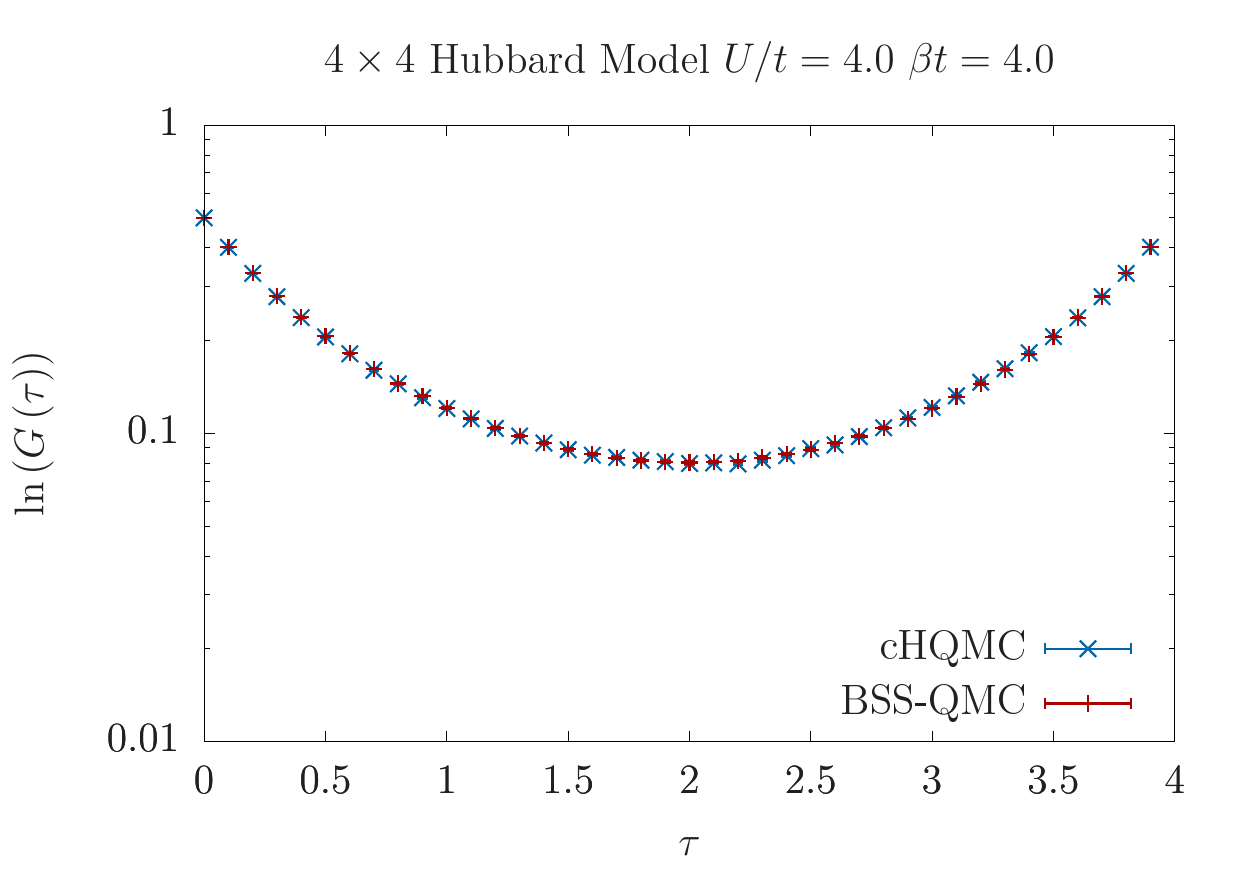}
 \caption{\label{fig_compare2} Single particle Green's function, calculated with the complex HQMC algorithm and compared to  the results of a BSS-QMC run \cite{2017ALF} for the same parameters as Fig.~\ref{fig_compare}.}
\end{figure}
\begin{figure}
 \includegraphics[width=.45\textwidth]{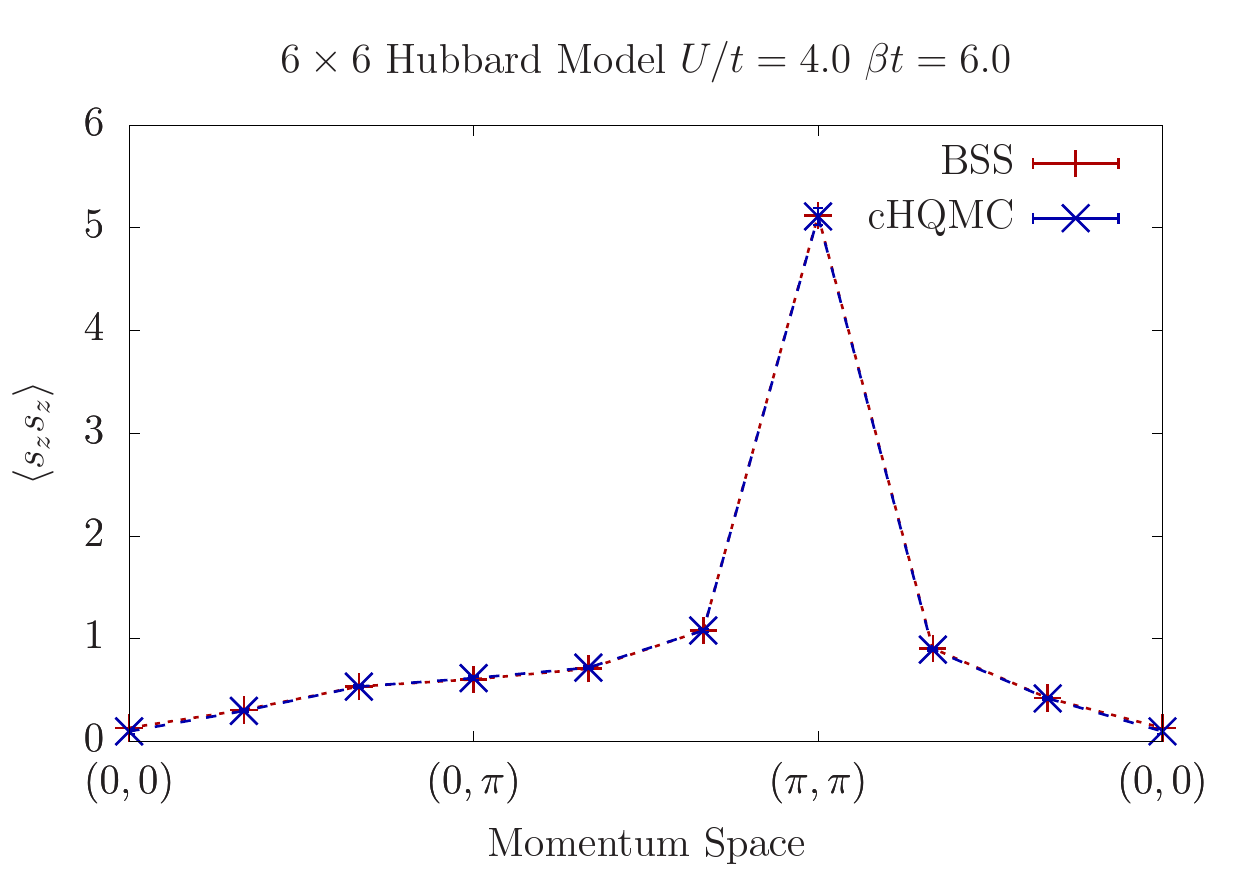}
 \caption{\label{fig_spsp} HQMC results for the  spin-spin-correlation functions in momentum space. Good agreement with benchmarks results from the BSS  runs \cite{2017ALF} is obtained.}
\end{figure}

If we now recalculate the single particle Green's function with the same parameters, where the real version of the algorithm previously failed, we observe consistent results, as shown in Fig.~\ref{fig_compare2}.
Beside the single particle Green's function we can also calculate higher observables like the spin-spin correlation function of a system, as shown in Fig.~\ref{fig_spsp}.
Wick's theorem allows us to decompose many particle Green's functions into expressions involving only single particle Green's functions.
Because we calculate the Green's functions only stochastically we get additional noise for many particle Green's functions and hence have to generate more samples.

\begin{figure}
 \includegraphics[width=.45\textwidth]{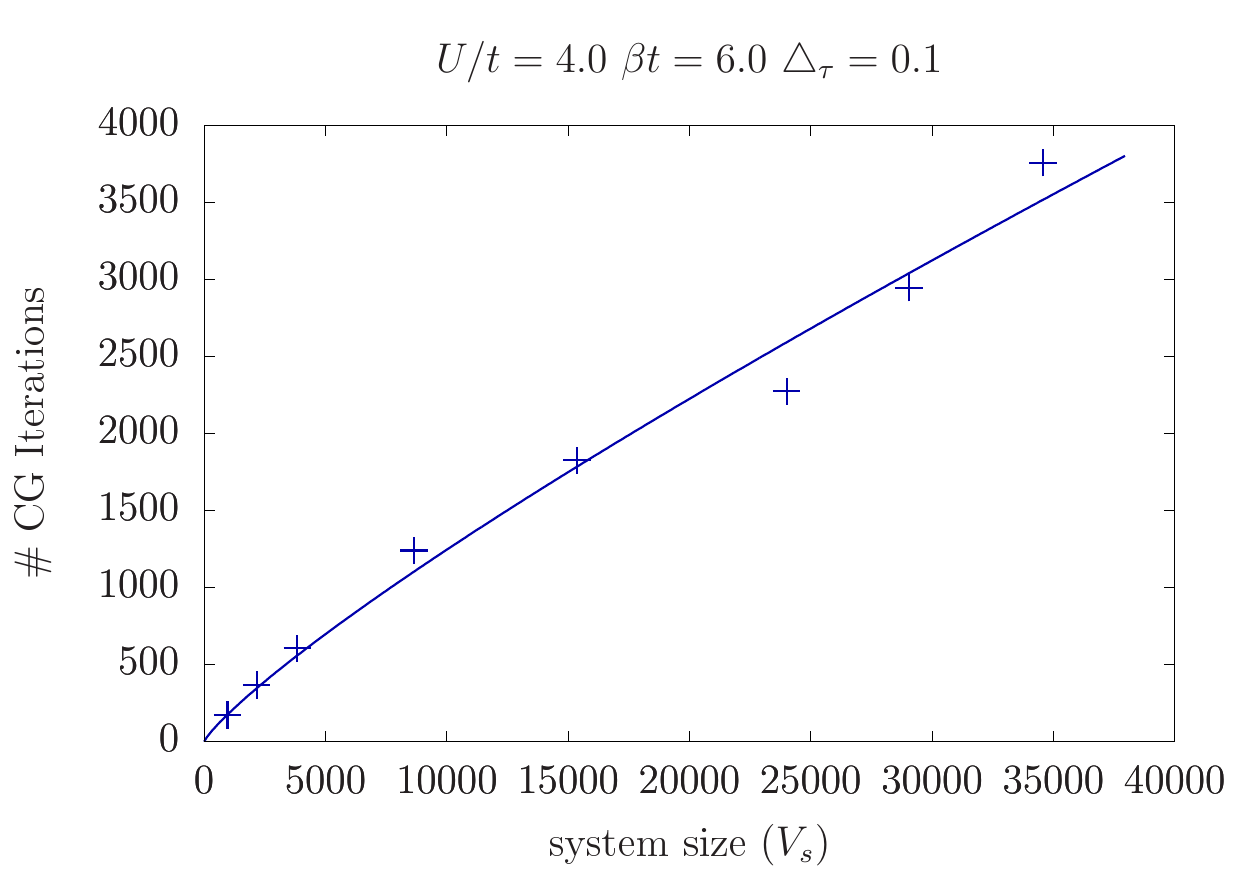}
 \caption{\label{CGitBSS} Average number of CG iterations needed, to solve a system of linear equations,  as a function of  the Euclidean system size as given by the number of Trotter slices times the lattice size ($N_{\tau}V_{s}$). For a two-dimensional system at fixed temperature, the plot shows an approximately linear behavior as a function of the lattice size.}
\end{figure}

To analyze the runtime of the algorithm  we carried out simulations in one, two and three dimensions  at different inverse temperatures. 
Generically,  and contrary to our aspirations,   the runtime  does not seem to scale favorably with system size and inverse temperature. 
There  are many reasons for this. 
The first is the observation that the number of iterations required for the conjugate gradient method  to converge  grows nearly linearly with the Euclidean size of the system (Fig.~\ref{CGitBSS}).
Another reason is that to keep the acceptance high, we have to rescale the artificial Leapfrog time step as  $\triangle t \propto V_{s}^{-1}\beta^{-1} $.
Combined with the linear scaling that every vector operation needs we achieved a scaling of approximately
\begin{equation}
 \left(L^{d}\beta\right)^{2.5-3.0}.
\end{equation}

\subsection{Comparison between cHQMC and BSS for the Hubbard model \label{compareHQMCBSS}}
\begin{figure}
 \includegraphics[width=.45\textwidth]{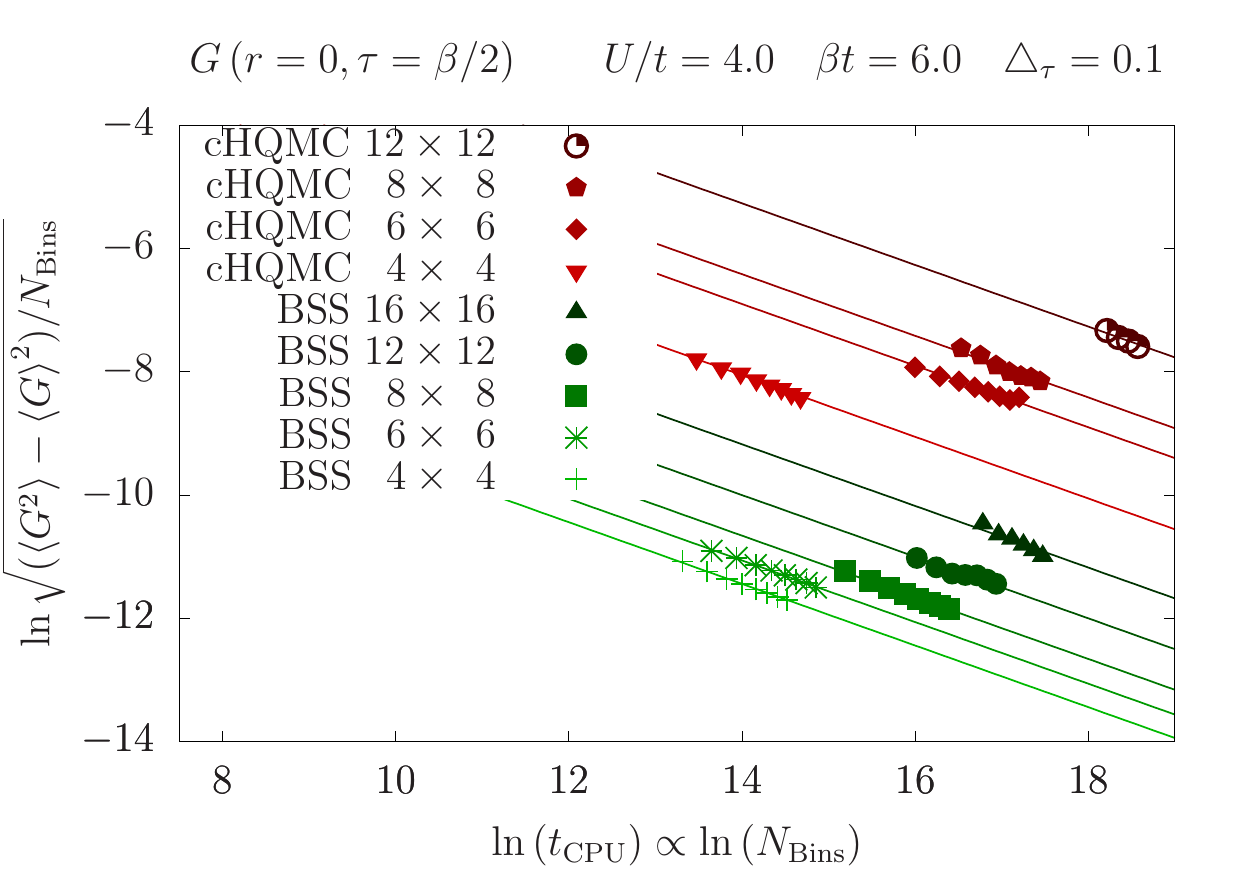}
 \caption{Comparison of the BSS and cHQMC.  Here we compare  error bars $\sigma$ as a function of computing time  for several system sizes.\label{fig_Compare_BSS_HQMC}} 
\end{figure}

Finally we have to compare  the efficiency of the cHQMC  with other simulation techniques.
The BSS-QMC method as implemented in the ALF software package \cite{2017ALF} is a well established and optimized algorithm to simulate half-filled Hubbard systems at finite temperature.
Both methods have the same Trotter-decomposition induced, systematic error which predestines them for a comparison.
Besides properties such as  required memory  and effective scaling for a single step,  fluctuations due to the statistical nature of the approach is a key property of every Monte Carlo method.
Owing to the central limit theorem, error bars decrease as the  square root of the number of measured samples, i.e., computing time. Fluctuations correspond to the  prefactor of this behavior.
Figure ~\ref{fig_Compare_BSS_HQMC} shows the decay of stochastic errors for both methods for several system sizes. Both methods show the expected behavior as a function of computational time.     However, for given computational time, the BSS method achieves  much higher precision than the cHQMC.
For this comparison we have carefully chosen the parameter set.
It is known that the spin correlation length of the half-filled Hubbard model grows exponentially with inverse  temperature.
The temperature regime where this scaling is valid corresponds to the so-called renormalized classical regime \cite{Chakravarty88}.
Our choice of parameters, $U/t=4$ and  $\beta t = 6$  places us in this regime \cite{White89},  which can be considered as {\it hard} for Monte Carlo simulations.   

\section{HQMC and the SSH model}\label{sec:HQMCSSH}

The question we would like to pose in this section is if  there exists a class of models in the solid state where the HQMC  is the method of choice.    We will argue that the Su-Schrieffer-Heeger (SSH) model \cite{SSHmodel}  describing the electron-phonon problem could lie in this class.  An obvious difficulty for  the HQMC method is singularities in the effective action trigger by the sign change  of the determinant in a given spin sector.  Using a Majorana representation,  one will show that  the determinant in a given spin sector is always positive semidefinite for the SSH model at the particle-hole symmetric point. 
 Furthermore, other Monte Carlo methods face issues for this electron-phonon problem. In the  CT-INT approach  \cite{Rubtsov04,Assaad07,Assaad14_rev}  the phonon degrees of freedom  are integrated out but in spacial dimensions larger than unity  and away from the antiadiabatic limit this generates a negative sign problem.   In the  BSS algorithm where phonons  degrees of freedom are sampled,  one can foresee that  local moves will lead to large autocorrelation times \cite{LangHohenadler2008}  such that global updating schemes such as Langevin dynamics or hybrid Monte Carlo should perform better.

\subsection{HQMC formulation for the SSH model}
The SSH Hamiltonian is given by
\begin{equation}
\hat{H}=\hat{H}_{\mathrm{el}}+\hat{H}_{\mathrm{ph}}+\hat{H}_{\mathrm{ep}}.
\end{equation}
Here, 
\begin{equation}
 \hat{H}_{\mathrm{el}}=-t\sum_{\left\langle i,j\right\rangle ,\sigma}\left(\hat{c}_{i,\sigma}^{\dagger}\hat{c}_{j,\sigma}+\hat{c}_{j,\sigma}^{\dagger}\hat{c}_{i,\sigma}\right)
\end{equation}
is the kinetic energy and $\left\langle i,j\right\rangle$ denotes the nearest neighbors of a square lattice.  Harmonic oscillators on links account for the lattice vibrations,
\begin{equation}
 \hat{H}_{\mathrm{ph}}=\sum_{\left\langle i,j\right\rangle}\left[\frac{\hat{P}_{\left\langle i,j\right\rangle}^{2}}{2m}+\frac{k}{2}\hat{Q}_{\left\langle i,j\right\rangle}^{2}\right],
\end{equation}
with  $\hat{P}, \hat{Q}$ being the conjugate momentum and position  operators. 
The  electron-phonon coupling  leads to a modulation of the hopping matrix element:
\begin{equation}
 \hat{H}_{\mathrm{ep}}=g\sum_{\left\langle i,j\right\rangle ,\sigma}\hat{Q}_{\left\langle i,j\right\rangle }\left(\hat{c}_{i,\sigma}^{\dagger}\hat{c}_{j,\sigma}+\hat{c}_{j,\sigma}^{\dagger}\hat{c}_{i,\sigma}\right)
\end{equation}
with a coupling strength $g$.
To simplify the notation we label bond indices as
\begin{equation}
  b := \left\langle i,j\right\rangle,
\end{equation}
and introduce the bond hopping as
\begin{equation}
  \hat{K}_{b} :=\left(\hat{c}_{i}^{\dagger}\hat{c}_{j}+\hat{c}_{j}^{\dagger}\hat{c}_{i}\right).
\end{equation}
To formulate the path integral, we will chose  a real space basis: 
\begin{equation}
 \hat {Q}_{b}\left|x_{b}\right\rangle =x_{b}\left|x_{b}\right\rangle,
\end{equation} 
such that the partition function reads,
\begin{equation}
 Z=\mathrm{Tr}\int\prod_{b}\mathrm{d}x_{b}\left\langle x_{b}\right|\mathrm{e}^{-\beta \hat{H}}\left|x_{b}\right\rangle,
\end{equation}
where the trace runs over the fermionic degrees of freedom. 
 The standard real space path integral and integration over the fermionic degrees of freedom yields the result:
 \begin{equation}
 Z=\int\prod_{b,\tau}\mathrm{d}x_{b,\tau}\:\mathrm{e}^{-S_{0}\left(x\right)}\left[\det M\left(x \right)\right]^{N_{\mathrm{col}}}, \label{Zssh}
\end{equation}
which is similar to the representation of the Hubbard model in Eq.~\eqref{eq_introduceDets}.     Being a phonon field, as opposed to a Hubbard-Stratonovich one,  $x_{b,\tau}$  has 
a bare imaginary time  dynamics given by the action
\begin{equation}
 S_{0}\left(x\right):=\triangle_{\tau}\sum_{b,\tau}\left(\frac{m}{2}\left[\frac{x_{b,\tau+1}-x_{b,\tau}}{\triangle_{\tau}}\right]^{2}+\frac{k}{2}x_{b,\tau}^{2}\right)
\end{equation}
of the harmonic oscillator. 
In the above, we have considered a model with   $N_{\mathrm{col}} $  spin components corresponding to an SU($N_{\mathrm{col}}$) symmetric model.
 The $ M\left(x\right)$ matrix has the same block structure as in Eq.~\eqref{B_Block} with
\begin{equation}
   B_{\tau}=\exp\left\{ -\triangle_{\tau}\sum_{b}\left(gx_{b,\tau}-t \right)K_{b}\right\}  . \label{Bmajo}
\end{equation}
To show the absence of singularities in the action one  should   demonstrate that  $\det M \left(x \right) >  C$ with  $C$ a finite positive constant.  Here we can only show a weaker statement,   namely that  $ \det M \left( x \right)  \ge 0 \; \forall x $.
Considering only one color degree of freedom,  and the Majorana representation on sublattices $A$ and $B$,  
\begin{equation}
 \begin{split}
 i  \in  A:&\quad\hat{\gamma}_{i,1}=\left(\hat{c}_{i}+\hat{c}_{i}^{\dagger}\right)\:\qquad\hat{\gamma}_{i,2}=-\mathrm{i}\left(\hat{c}_{i}-\hat{c}_{i}^{\dagger}\right)\\
i  \in  B:&\quad\hat{\gamma}_{i,1}=-\mathrm{i}\left(\hat{c}_{i}-\hat{c}_{i}^{\dagger}\right)\quad\hat{\gamma}_{i,2}=-\left(\hat{c}_{i}+\hat{c}_{i}^{\dagger}\right)
\end{split}
\end{equation}
the exponent in Eq.~\eqref{Bmajo} reads
\begin{align}
 \sum_{b}\left(-t+gx_{b,\tau}\right)&\left(\hat{c}_{i}^{\dagger}\hat{c}_{j}+\hat{c}_{j}^{\dagger}\hat{c}_{i}\right)\\
 &=\frac{\mathrm{i}}{2}\sum_{b,n}\left(-t+gx_{b,\tau}\right)\hat{\gamma}_{i,n}\hat{\gamma}_{j,n}. \nonumber
\end{align}
The above result depends upon the fact that the hopping matrix element is real and that hopping occurs only between different  sublattices. 
Thereby, the trace factorizes 
\begin{align}
 \mathrm{Tr}\prod_{\tau=1}^{N_{\tau}}&\mathrm{e}^{-\triangle_{\tau}\sum_{b}\left(-t+gx_{b,\tau}\right)\left(\hat{c}_{i}^{\dagger}\hat{c}_{j}+\hat{c}_{j}^{\dagger}\hat{c}_{i}\right)} \\
 &=\left[\mathrm{Tr}\prod_{\tau=1}^{N_{\tau}}\mathrm{e}^{-\mathrm{i}\frac{\triangle_{\tau}}{4}\sum_{b}\left(-t+gx_{b,\tau}\right)\hat{\gamma}_{i}\hat{\gamma}_{j}}\right]^{2}, \nonumber
\end{align}
and one can show  that the trace  over a  one Majorana mode is a real quantity\cite{Yao14a}.   Thereby, $ \det M \left( x \right)  \ge 0 \; \forall x $.

Going on to implement a HQMC method for the SSH model, we define 
\begin{equation}
\tilde{x}_{b,\tau}:=\sqrt{\frac{k}{2}}x_{b,\tau}\qquad
\omega_{0}^{2}:=\frac{k}{m}\qquad
\tilde{g}:=\sqrt{\frac{2}{k}}g
\end{equation}
and include the canonical conjugate momentum and pseudofermions to obtain:
\begin{widetext}
\begin{equation}
 Z=\int\left[\mathrm{\delta}x\,\delta p\,\delta\phi_{\sigma}\right]\exp \Biggl\{ -\underbrace{\triangle_{\tau}\left[\sum_{b}\left(\omega_{0}^{-2}\left[\frac{\tilde{x}_{b,\tau+1}-\tilde{x}_{b,\tau}}{\triangle_{\tau}}\right]^{2}+\tilde{x}_{b,\tau}^{2}\right)\right]}_{=S_{0}\left(\tilde{x}\right)}-\sum_{\sigma=N_{\mathrm{col}}}\phi_{\sigma}^{\mathrm{T}}\left(M^{\mathrm{T}}M\right)^{-1}\phi_{\sigma}-\sum_{b}p_{b,\tau}^{2}\Biggr\} \label{elphoZ}
\end{equation}
\end{widetext}

Henceforth, everything is similar to implementation for the Hubbard model.  After generating random numbers for $p$ and $\phi_{\sigma}$ fields a Leapfrog run updates the auxiliary field, followed by measurements of observables before the loop starts again.
A CG method   is put to use to solve the system of linear equations.

\subsection{Proof of Concept}

\begin{figure}

 \includegraphics[width=.45\textwidth]{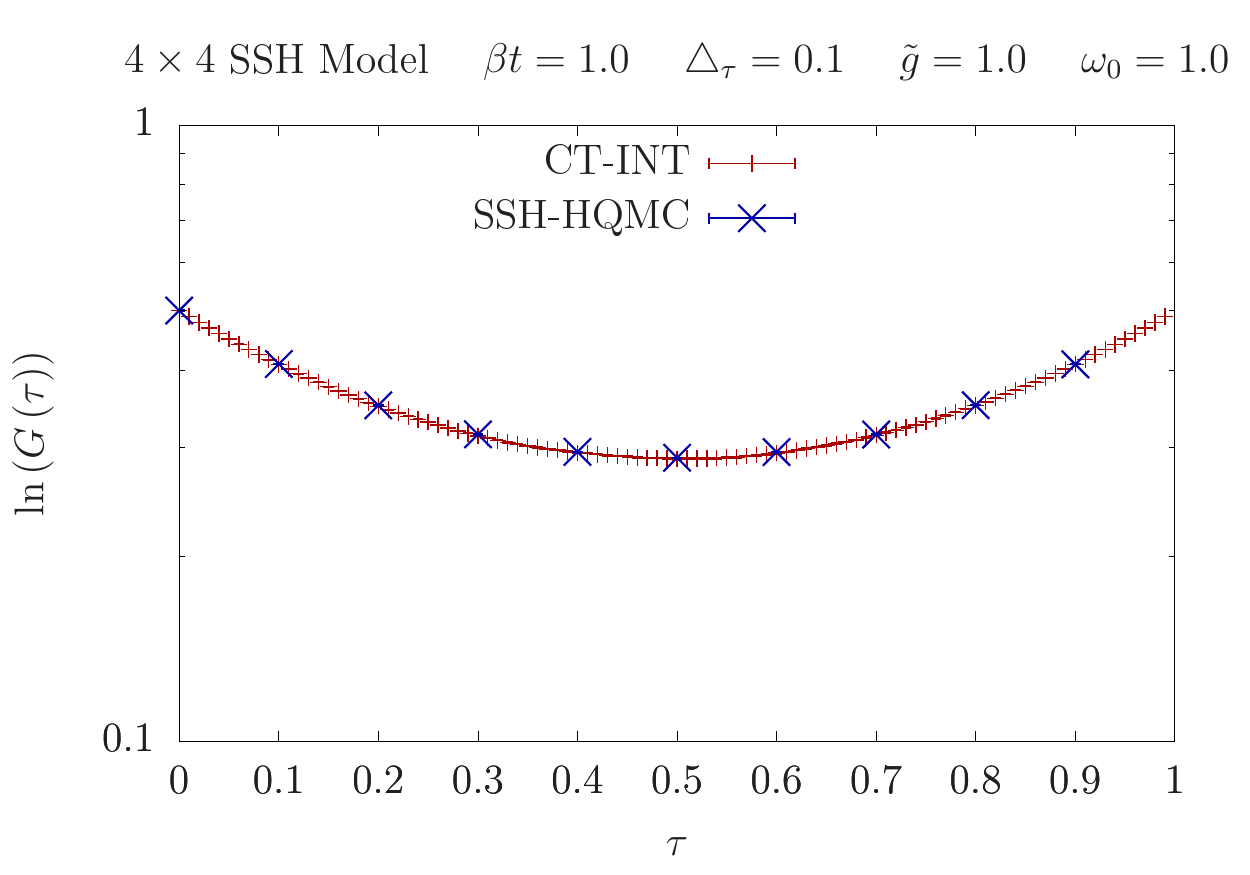}
 \caption{HQMC and CT-INT results for the local Green function  for SSH model.
 \label{SSH1.fig}}
\end{figure}

\begin{figure}
 \includegraphics[width=.45\textwidth]{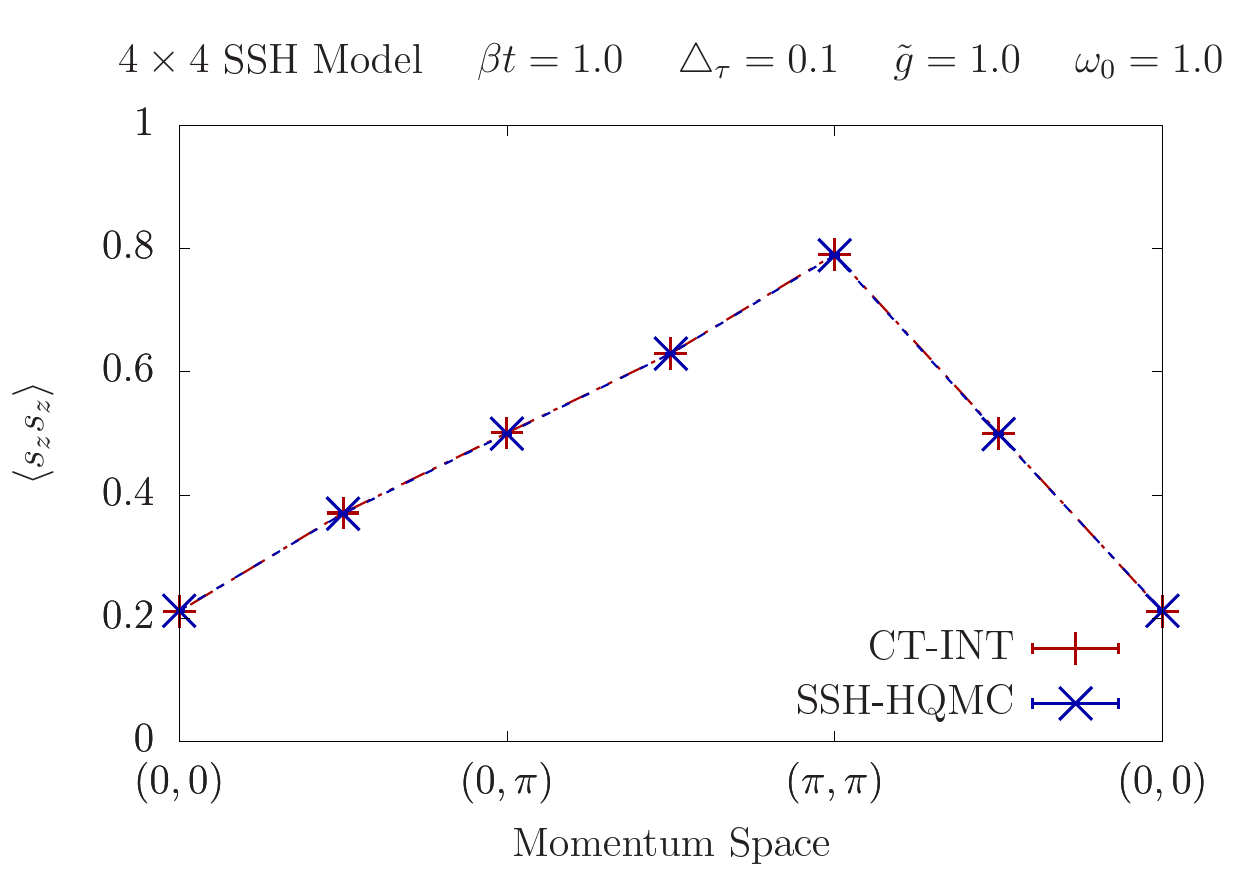}
 \caption{ HQMC and CT-INT spin-spin correlation functions in the momentum space for the SSH model.
  \label{SSH2.fig}}
\end{figure}

In Fig.~\ref{SSH1.fig}  and Fig.~\ref{SSH2.fig}  our HQMC  results for the SSH model are benchmarked against CT-INT simulations in a regime where the latter method does not suffer from a severe sign problem. As apparent, perfect agreement is obtained. 

\begin{figure}
 \includegraphics[width=.45\textwidth]{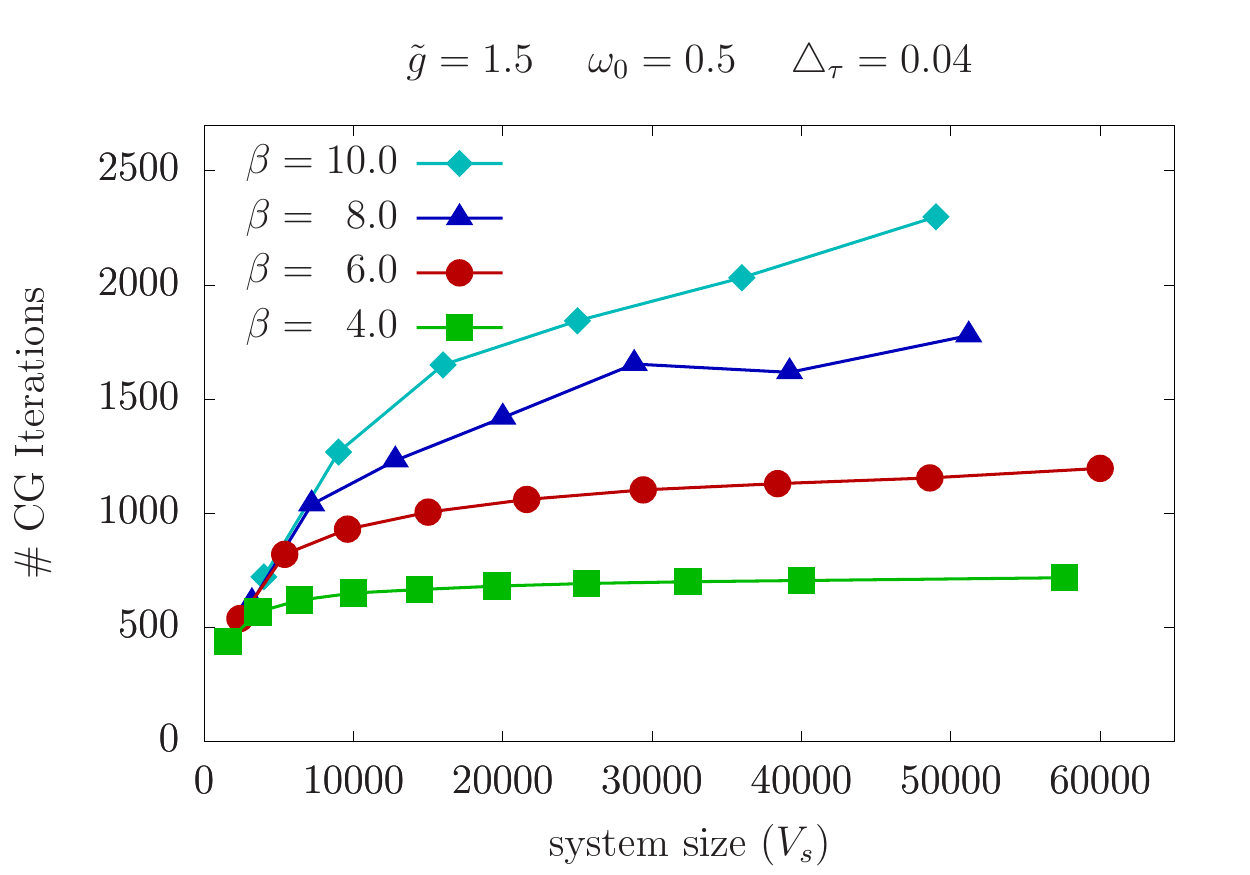}
 \caption{Number of CG iterations  required for a given  accuracy as functions of the lattice size at various temperatures.  
 At each temperature one can  observe  saturation. The lattice size at which this saturation kicks in is  temperature dependent. 
 Here we considered  $4\!\times\! 4$,$\;6\!\times\! 6$,$\;8\!\times\! 8$,$\;10\!\times\! 10$,$\;12\!\times\! 12$,$\;14\!\times\! 14$,$\;18\!\times\! 18$, $\;20\!\times\! 20$ and $\;24\!\times\! 24$  lattices.
 \label{SSH3.fig}}
\end{figure}

\begin{figure}
 \includegraphics[width=.45\textwidth]{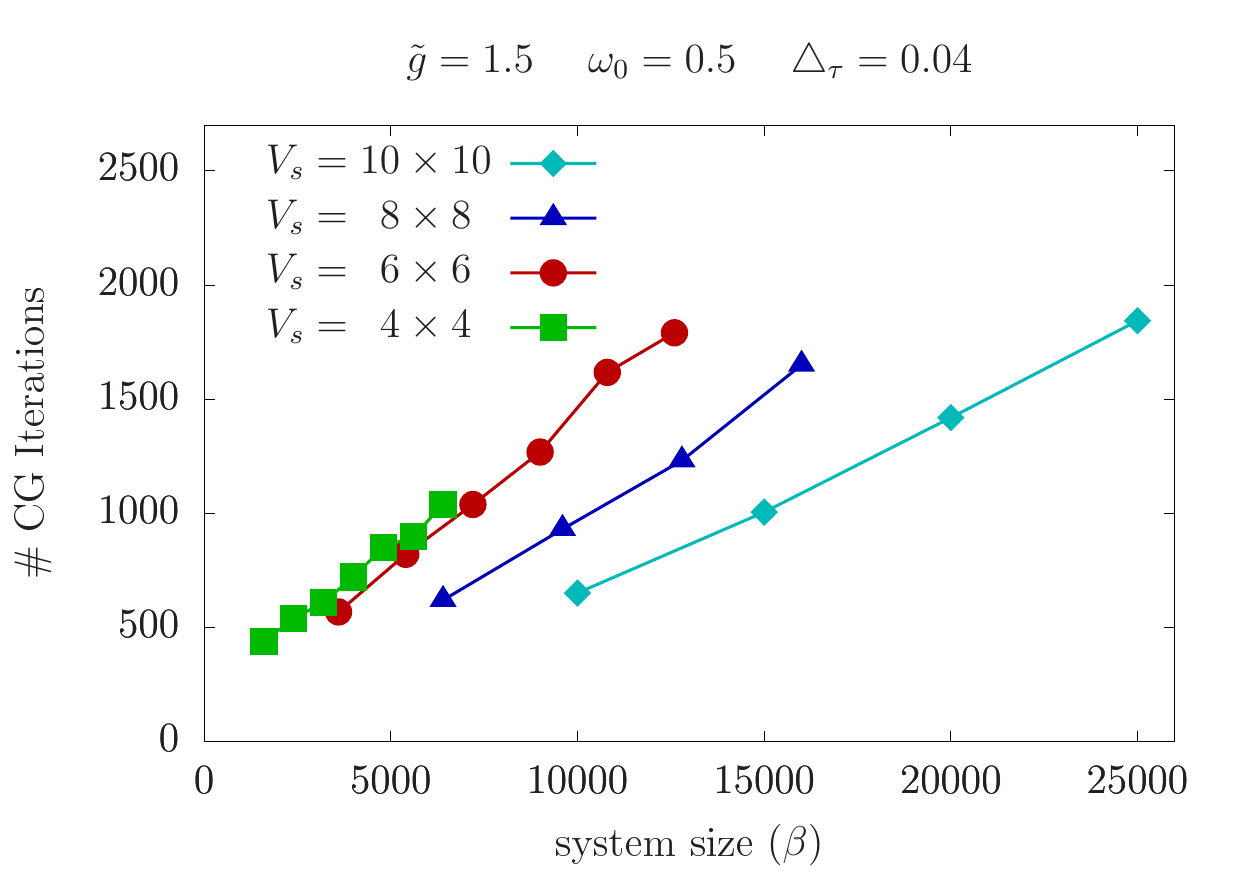}
 \caption{ Number CG  iterations as a function  of inverse temperature at different lattice sizes.   In this case, no saturation is observed  in the considered  temperature range.   Here we consider   $\beta$  =     $4.0$, $6.0$, $8.0$ and $10.0$.
 \label{SSH4.fig}}
\end{figure}

The scaling behavior of the HQMC method for the SSH model  seems more favorable than for the  Hubbard model.  We observe a much weaker dependence of the acceptance on the Leapfrog step size. For the SSH model a correction of approximately $\triangle t \propto V_{s}^{-0.25}\beta^{-0.25}$  was sufficient in most cases.  Note that for the Hubbard model, the time step had to be scaled as 
  $\triangle t \propto V_{s}^{-1}\beta^{-1}$.  We believe that this reflects the fact  that  singularities in the effective action are not an issue. 

Figure ~\ref{SSH3.fig} and Fig.~\ref{SSH4.fig}  plot the number of CG steps required  during the simulation so as to achieve the desired accuracy.   This result, alongside the favorable  Leapfrog step scaling,  gives an  estimate of the numerical effort: 
\begin{equation}
\left( L^{d}\right)^{1.25-1.5}\beta^{2.25-2.5}.
\end{equation}
It is interesting to  note the asymmetry in the temporal and spatial directions (see Figs.~\ref{SSH3.fig} and \ref{SSH4.fig}).   A possible understanding of this is in terms of  a dynamical exponent greater than unity that renders the characteristic length scales longer along the imaginary time  than in real space.

\section{Conclusion}
\label{Sec:conclusions}

By design, the HQMC algorithm  has the potential of  solving sign free fermion problems  with a numerical effort scaling linearly with  the Euclidean system size, $V_s \beta $.    The key ideas leading to this statement are the  following.
\begin{itemize}
\item Stochastic sampling of the fermion determinant.  
\item Global molecular dynamics updating schemes.
\item Stochastic sampling of the single particle Green functions   required for the computation of observables. 
\end{itemize}
Remarkably  the above  relies solely on  the solution of Eq.~\eqref{eq_CGinit},  $  M^{\mathrm{T}}M X = \phi $, and since $M$   contains  only order $V_s \beta $ nonvanishing matrix elements one 
can hope for linear scaling in  $V_s \beta $. 

For the Hubbard model at half filling, the major issues are the zeros of determinant of the fermion matrix $M$. Even though the  total weight is positive,  the sign of the  determinant in a single spin sector at \textit{ low }  temperatures fluctuates strongly.   We showed that this leads to  ergodicity issues since  the zeros split the configuration space in  regions separated by infinite potential barriers that cannot be overcome  with molecular dynamics.    To circumvent this problem we have complexified the  Hubbard-Stratonovich transformation so as to be able to \textit{ go around  }  these singularities.  Our main result, however, is that in comparison to the generic BSS algorithm as implemented in the ALF project  \cite{2017ALF}, the complexified HQMC for the Hubbard model  remains orders  of magnitude slower.    This statement relies on the following observations. (i)  Fluctuations  of standard observables are much larger. (ii) The roughness of the  potential landscape  renders an efficient implementation of the molecular dynamics challenging. In other words, the forces are hard to compute and the CG   approach fails to converge  in a number of iterations that scale with the Euclidean volume.  These statements are based on a  simple implementation of the  HQMC, and much progress has  been made to solve the above issues.   More efficient algorithms for the Hubbard model could be based on various strategies.  
One can  modify  the  action with for instance symmetry breaking terms  so as to avoid the aforementioned  singularities stemming from zero eigenvalues of the fermion  matrix.  This step can be combined with  rational HQMC \cite{Clark07} or mass preconditioning measures \cite{Hasenbusch01}  both aimed at reducing the condition number of the Fermion matrix.    Furthermore, other choices of the Hubbard-Stratonovich transformation, in particular compact versions \cite{Dlee08}, could provide a speedup.     
Another route is to bias the classical Hamiltonian  used for the molecular dynamics so as to  efficiently cope with the unbounded forces stemming from the zeros of fermion determinant.  This bias will be not lead to systematic errors due to  the Metropolis acception/rejection step at the  end of the Molecular dynamics integration. However the issue will be to ensure good acceptance. Such strategies have been implemented in the realm of overlap fermions  \cite{Neuberger98,KaplanB92,Chandrasekharan04,Cundy09,Fodor05}.

Since the sign changes  of the  fermion determinant  in a given spin sector render  the implementation  of the HQMC hard, one will  invariably search for interesting models where   the  determinant remains  positive for all field configurations.  Using recent insights from the Majorana representation \cite{Yao14a}  to classify sign free  Hamiltonians, one will show that the SSH model on a bipartite  lattice at half filling falls in this class.    This model describes  the  salient physics of  the electron-phonon interaction and is solvable in one dimension with the CT-INT approach \cite{Weber15}. In higher dimensions, its  phase diagram remains  illusive: A simple sign free implementation of the BSS will suffer from  long autocorrelation times whereas the  CT-INT approach in which the phonons are integrated out turns out to  suffer from a sign problem away from the antiadiabatic limit.   Our preliminary results   in solving this model with the HQMC   are very promising.    Away from half filling  and/or away from the   particle-hole symmetric point, the model for  an even number of fermion flavors  does  not suffer from the negative sign problem and can be simulated with the HQMC. In this case however,   the sign of the  determinant  for a single flavor can start  fluctuating and  thereby reduce the efficiency of the HQMC.  It is however unclear to what extent  the efficiency of the algorithm will    suffer  away from the particle symmetric point.     Further work in this direction is presently in progress.

\begin{acknowledgments}

We are very indebted to Martin Hohenadler for providing the SSH test data and for fruitful discussions on numerous occasions.
We acknowledge fruitful discussions with 
N. Brambilla,  
P. Buividovich, 
M. Hasenbusch,
A. Kennedy,  
M. K\"orner,
D. Schaich, 
L. Smekal, 
D. Smits, and 
J. Thies from the ESSEX project (\url{https://blogs.fau.de/essex/}), 
M. Ulybyshev,  
J. Weber, 
 and finally KONWIHR (http://www.konwihr.uni-erlangen.de)  for algorithmic support. 
The authors gratefully acknowledge the Gauss Centre for Supercomputing e.V. (\url{www.gauss-centre.eu}) for funding this project by providing computing time on the GCS Supercomputer SuperMUC at Leibniz Supercomputing Centre (LRZ, \url{www.lrz.de}).
We thank the DFG for funding through the SFB 1170 “Tocotronics” under the Grant No. C01 (F.F.A. and S.B.) as well as Z03 (F.G.).
\end{acknowledgments}

%

\end{document}